\documentclass[acmsmall]{acmart}

\AtBeginDocument{%
  }

\setcopyright{acmlicensed}
\copyrightyear{2026}
\acmYear{2026}
\acmDOI{XXXXXXX.XXXXXXX}

\acmISBN{978-1-4503-XXXX-X/2026/}

\usepackage{multirow}
\usepackage{booktabs}
 
\usepackage{tikz}
\usepackage{circuitikz}

\usepackage[most]{tcolorbox}
\usepackage{ulem}

\begin{document}
\newcommand{\Mas}[1]{{\color{purple} (Mas: #1)}}
\newcommand{\xh}[1]{{\color{magenta} (Xi: #1)}}

\newcommand{\highlight}[1]{{\color{blue} (Notes: #1)}}

\newcommand{\paratitle}[1]{\paragraph{\textbf{#1.}}}
\title[Adopt a PET! An Exploration of PETs, Policy, and Practicalities for Industry in Canada]{Adopt a PET! An Exploration of PETs, Policy, and Practicalities for Industry in Canada}


\author{Masoumeh Shafieinejad}
\affiliation{%
  \institution{Vector Institute}
  \country{Canada}}
\email{masoumeh@vectorinstitute.ai}

\author{Xi He}
\affiliation{%
 \institution{Vector Inst. \& Univesity of Waterloo}
 \country{Canada}}
\email{xi.he@waterloo.ca}

\author{Bailey Kacsmar}
\affiliation{%
  \institution{Amii \& University of Alberta}
  \country{Canada}}
\email{kacsmar@ualberta.ca}


\renewcommand{\shortauthors}{Shafieinejad et al.}

\begin{abstract}
Privacy is an instance of a social norm formed through legal, technical, and cultural dimensions. Institutions such as regulators, industry, and researchers act as societal agents that both influence and respond to evolving norms. Attempts to promote privacy are often ineffective unless they account for this complexity and the dynamic interactions among these actors.
Privacy enhancing technologies (PETs) are technical solutions for privacy issues that enable collaborative data analysis, allowing for the development of solutions that benefit society, all while ensuring the privacy of individuals whose data is being used. However, despite increased privacy challenges and a corresponding increase in new regulations being proposed by governments across the globe, a low adoption rate of PETs persists. In this work, we investigate the factors influencing industry's decision-making processes around PETs adoption as well as the extent to which privacy regulations inspire such adoption. 
We conducted a qualitative survey study with 22 industry participants from across Canada to investigate how businesses in Canada make decisions to adopt novel technologies and how new privacy regulations impact their business processes.  
Informed by the results of our analysis, we make recommendations for industry, researchers, and policymakers on how to support what each of them seeks from the other when attempting to improve digital privacy protections. 
By advancing our understanding of what challenges the industry faces, we increase the effectiveness of future privacy research that aims to help overcome these issues.

\end{abstract}


\received{X}
\received[revised]{X}
\received[accepted]{X}

\maketitle

\section{Introduction}\label{sect:intro}

Societal demand for privacy is influencing law-makers, resulting in ongoing introductions of new privacy regulations globally~\cite{ccpa2018,gdpr2016,billc27,edwards2021eu}. 
Most Canadian companies are moderately aware of their responsibilities under Canada’s privacy laws and have taken steps to ensure they comply with these laws according to the most recent survey (2023-2024) of Canadian businesses~\cite{phoenixspi2024survey}.
However, while there has been a lot of effort in privacy enhancing technologies (PETs) research to develop deployable privacy-enhancing tools that aid organizations in their privacy preservation approaches, it is unclear whether additional regulations will motivate increased usage of these tools by industry. 
The divide between research and practice is a prolific one, and it is known that technical developments and the social components of their use in practice can have a great divide~\cite{ackerman2000intellectual}. 

The implications of social contexts for knowledge sharing is an unavoidable first hurdle to bridging the divide before the adoption of novel technologies~\cite{ackerman2003sharing,ackerman2013sharing}. In high risk domains, such as emergency rooms, it has been established that better systems emerge when consideration is given to how systems get to those who need them and how to meet those users needs~\cite{normark2005local}. 
Thus, researchers need to understand how industry approaches expertise collection if we are going to improve and resolve the gaps and issues originating from different expectations on expertise communication and gathering strategies~\cite{pipek2012bridging,halverson2004behind}.

We investigate the relationship of technical developments of PETs to the social practice of deployment of PETs in conjunction with how emergent policies and regulations are influencing the adoption or avoidance of such technologies. In doing so we observe PETs to be a notable example of the interconnected nature of policy, practice and design for systems aimed at social-technical problems like digital privacy~\cite{Jackson2014} -- which is what PETs tries to protect.
Specifically, we address the following research questions in the context of Canada's industry and privacy regulations.
\paratitle{Research questions}
\begin{itemize}
    \item \textbf{RQ1:} How does regulation influence industry adoption of PETs? 
    \item \textbf{RQ2:} What barriers does industry perceive as preventing them from wanting to or choosing to adopt PETs? 
    \item \textbf{RQ3:} What are industry's common practices in regard to sensitive data and PETs?  
    \item \textbf{RQ4:} 
    How does industry determine whether they are compliant with relevant laws, regulations, or policies? 
\end{itemize}adopt a pet arxiv 

We highlight the social challenges expressed by industry as well as propose ways to overcome these issues that contribute to such divides. We focus on the identification of existing or perceived impediments to adopting these technologies and complying with regulations to better understand how the industry-research relationship can be improved. 
To address our research questions we employ a qualitative survey study with 22 participants. We employ thematic analysis over the collected responses to address our research questions and make the following contributions. (i) We identify insights from industry in regard to how privacy regulations impact their business and their current practices for handling sensitive data. (ii) From our respondents' reported processes, we derive the components of the decision-making process for PETs adoption and its inclusion of both personnel inside and outside of the organizations. (iii) We determine that challenges that impede the adoption of PETs include compatibility with existing systems, feature requirements, lack of clarity for how they fulfill regulatory requirements, and the broader socio-economic system of industry.




\paratitle{Organization} The remainder of this paper is organized as follows. We present background on PETs and Canada in Section~\ref{sect:background} followed by the relevant related work in Section~\ref{sect:relwork}, and our methodology is included in Section~\ref{sect:methodology}. We organize our results within each research question. Section~\ref{sect:RQ1} is our results for RQ1, Section~\ref{sect:RQ2} is our results for RQ2, Section~\ref{sect:RQ3} is our results for RQ3, and Section~\ref{sect:RQ4} is our results for RQ4. We provide additional discussion on the key results and impacts of our work in Section~\ref{sect:discussion} before ending with our conclusion in Section~\ref{sect:conclusion}.

\section{Background}\label{sect:background}


\paratitle{What is and is not a PET according to researchers}
Privacy-enhancing technologies (PETs) are technical solutions that aim at protecting privacy, a complex and evolving concept. These technologies address diverse aspects of privacy, as illustrated by Heurix et al.~\cite{HEURIX20151} in their 2015 taxonomy study. Their framework categorizes PETs across seven dimensions, including scenarios (e.g., untrusted client/server), aspect (identity, content, behavior), aim (indistinguishability, unlinkability, deniability, confidentiality), foundations (security model, cryptography), data (stored, transmitted, processed), trusted-third party, and reversibility. Their evaluation covers a wide range of PETs, from 1-n oblivious transfer (OT)~\cite{10.5555/36664.36681} in the 1980s to anonymization techniques~\cite{kanonymity02,ldiversity07} developed in the 2000s. While many of these PETs have seen practical adoption, their effectiveness can be challenged by emerging attacks and evolving system environments. For instance, k-anonymity~\cite{kanonymity02}, initially considered a promising technique for hiding personal identity by masking or generalizing (quasi)-identifiable information, was then shown to be vulnerable to new attacks~\cite{ldiversity07}, prompting the development of stronger PETs like differential privacy (DP)~\cite{Dwork2006DP}. Similarly, while access control (AC) as a PET~\cite{petshandbook2003} has been widely adopted by industry for decades,  studies have found that the deployed AC mechanisms are not necessarily compatible with privacy regulations like GDPR~\cite{gdprac22}, and there are challenges in unifying different AC models and techniques for enforcement purposes, particularly in federated environments~\cite{ac18}.  

This work considers three types of PETs based on their industrial deployment and research maturity. The first type includes 
PETs that are widely adopted in the industry but may not represent the state-of-the-art (SOTA). Examples include k-anonymization for sensitive data release and access control for managing sensitive databases. The second category comprises PETs with widespread deployment and relatively few identified critical vulnerabilities, such as multi-factor authentication~\cite{10.1145/3440712} for online login,  SSL/TLS~\cite{10.1145/1298306.1298318} for data transmission, and AES encryption~\cite{HERON20098} for data storage.
Finally, the third category consists of SOTA PETs, which are actively researched but not yet broadly deployed in the industry. These include differential privacy~\cite{Dwork2006DP}, multi-party computation~\cite{mpc87}, oblivious RAMs~\cite{oram96,oram19}, homomorphic encryption~\cite{he15}, trusted execution environments (TEEs)~\cite{tee15}, federated learning~\cite{fl21}, and synthetic data~\cite{JPMorgan2024synthetic,NIST2018}. Our study focuses on understanding what challenges impair industry adoption of these emerging PETs.

While researchers often define PETs as specific algorithms or systems that achieve desired privacy properties, the industry commonly interprets the term more broadly as general practices and principles. These include privacy impact/risk assessments, privacy policy development, and privacy control implementation. Although these practices are essential for establishing privacy within industrial settings and can be significantly supported by PETs, they are not, in themselves, PETs. Rather, they are the frameworks and processes within which PETs can be effectively deployed.




\paratitle{Why do PETs experts think PETs are good for industry} PETs experts advocate for the adoption of PETs within the industry due to their multifaceted benefits. They believe PETs primarily empower organizations to achieve and maintain data protection and compliance with evolving privacy regulations through technical tools for anonymization, data minimization, and secure data processing~\cite{garrido2023lessons,birrell2024sok}. With increasing data breaches and privacy concerns, businesses prioritizing robust data protection gain a significant competitive advantage, thus fostering customer trust and loyalty~\cite{garrido2023lessons}. Furthermore, PETs, like differential privacy and secure multi-party computation, facilitate secure data sharing and collaborative analysis of sensitive data with both internal and with external partners in sectors such as healthcare, finance, and research~\cite{NIST2018,cummings2023centering}. Finally, technologies such as oblivious RAM and homomorphic encryption enable organizations to confidently outsource sensitive data and computations backed by strong security guarantees~\cite{oram19,ZHAOE201973}. 

\paratitle{Canadian privacy regulations} Personal Information Protection and Electronic Documents Act (PIPEDA)~\cite{pipeda2000} is Canada's federal privacy law for private-sector organizations. It sets out the ground rules for how businesses must handle personal information in the course of their commercial activity. The Canadian government proposed Bill C-27~\cite{billc27} in 2022 to enact the Consumer Privacy Protection Act, the Personal Information and Data Protection Tribunal Act and the Artificial Intelligence and Data Act (AIDA). While, the federal privacy law reforms and artificial intelligence regulation contained in the bill are put in an uncertain state following the prorogation of Parliament in January 2025, their potential impact on Canada's industry cannot be ignored~\cite{iapp2025billc27}. 
Further, each of Canada's provinces and territories has their own additional privacy regulations and either a commissioner or ombudsperson who is responsible for the privacy legislation in that region~\cite{provinces}.
Finally, we note that in Canada, the privacy regulations apply to all sectors and are not separated out by areas such as health having their own regulations. Rather, all sectors must protect that which is deemed to be potentially privacy-harmful regardless of whether the sector that initiated the data collection is a sector considered to be high-risk or privacy critical. 
\section{Related work}\label{sect:relwork}

In our work, we center the participants' views and experiences, rather than centering our own preconceived notions or those of the broader research community, to allow us to highlight the differences of interpretations of PETs terminologies (e.g., PETs itself, anonymization, best practices for privacy, etc.) between industry and researchers. 
Prior research mainly focuses on evaluating comprehension or expectations for a particular PET, such as differential privacy (DP), multiparty computation (MPC), or homomorphic encryption (HE). 
We provide an overview of related research that seeks to motivate and facilitate the deployment of PETs by examining their interaction with key actors and influencing factors, such as industry needs and regulatory requirements.

\paratitle{PETs deployment challenges and facilitators}
Numerous studies examine the challenges involved in deploying specific PETs within industry settings. As a promising technique that supports anonymity, Differential privacy (DP)~\cite{Dwork2006DP} has been the center of attention~\cite{cummings2023centering, Dwork2019DPinPractice,garrido2023lessons}. Notably, NIST produced a set of guidelines for evaluating DP guarantees~\cite{near2023guidelines}. Dwork et al. study deployment challenges by gathering DP experts' opinions in 2019~\cite{Dwork2019DPinPractice} and distill them into four main categories. Their findings were as well mirrored in the results of a study by Garrido et al. in 2023~\cite{garrido2023lessons} on industry participants who were not experts in DP. 
Several research evaluate or enhance the usability,  deployability, and communicative ease of PETs for real-world scenarios. 
For DP, they cover a wide range, including: user expectations of the DP mechanism ~\cite{cummings2021need} and their mental models of DP open-source libraries~\cite{song2024inherently}, the usability of DP tools~\cite{ngong2024evaluating}, the need for explanation of the gained privacy~\cite{nanayakkara2023chances} in user-suitable metaphors \cite{karegar2022exploring} or in risk communication format~\cite{franzen2022private}, and considerations for policy makers~\cite{nanayakkara2024consider}. The studies of user experience extends to other forms of private computation, such as multi-party computation and private query execution as well~\cite{kacsmar2023comprehension}. Qin et al.~\cite{Qin2019FromUsability} yield insights about the interplay between usability and security in multi-party computation (MPC) applications. Agrawal et al.~\cite{Agrawal2021Exploring} focus on the journey of a number of privacy-preserving computation (PPC) techniques -- namely, homomorphic encryption (HE), secure MPC, and DP -- from research into production code. They focus on specific application contexts; usability of the libraries and tools from a non-specialist developer’s perspective; and the explanation and governance challenges associated with these techniques. Since PETs development relies heavily on cryptography in several examples, we also include the challenges in cryptography's journey from research paper to products. Fischer et al.~\cite{Fischer2024Challenges} identify these challenges as: misunderstandings and miscommunication among stakeholders, unclear delineation of responsibilities, misaligned or conflicting incentives, and usability.

In contrast to the prior research, our work not only includes the technical motivators to PETs adoption, but also the strategic influences --  factors that determine whether PETs succeed or fail in real-world cost-benefit evaluations. We explore how industry perceives PETs, what drives their interest, and the obstacles that hinder adoption.


\paratitle{PETs and regulations} The introduction of privacy regulations has made evident changes in PETs research papers. Some discuss how their proposed technology aligns with privacy regulations. For example, synthetic data generation technologies are perceived to provide a practical replacement for the original sensitive data~\cite{NIST2018, JPMorgan2024synthetic}. Others draw inspiration from the tension between data privacy policies and socially beneficial analytics to develop usable, privacy-preserving systems -- Synq~\cite{Espiritu2024Synq} for instance, exemplifies this in computation over encrypted data. Additionally, some PETs research gives dedicated attention to the technology's relationship with regulations. Walsh et al.~\cite{Walsh2022Multi-Regulation} examine privacy laws and regulations that limit disclosure of personal data, and explore whether and how these restrictions apply when participants use cryptographically secure multi-party computation (MPC). Another body of work focuses on assessing the privacy promises of PETs~\cite{stadler2022synthetic, Giomi2023AUnified}. Giomi et al. introduce Anonymeter, a framework for quantifying privacy risk in synthetic data~\cite{Giomi2023AUnified}, equipped with attack-based evaluations for the singling out, linkability, and inference risks, which are the three key indicators of factual anonymization according to data protection regulations, such as GDPR.

As regulations increasingly mandate the deployment of privacy-preserving mechanisms -- and with the emergence of Canada's new privacy regulation (C-27) serving as a key inspiration to this work -- we included a focused investigation into how regulations influence PETs adoption.


\section{Methodology}\label{sect:methodology}

Our study is targeted at the needs, norms, and expectations of industry actors in Canada in regard to privacy regulation and privacy technologies. 
As the needs and expectations of industry actors in Canada regarding privacy technology adoption and the influence of regulation are not well known, we wanted to avoid prematurely narrowing the scope of permitted responses to something that would fit within a quantitative survey style. We also faced challenges with time availability and anonymity were we to do a focus group or interview style study. These challenges were communicated to us during our recruitment effort as being based in industry's information sharing restrictions for their employees. Thus, we chose to do an open-ended survey style to (i) ease access and effort for our intended participant pool while still (ii) affording freedom to our participants concerning how much or how little they shared with us. 
In the end, our survey facilitated a  non-presumptuous understanding of industry’s perspective on privacy, and allowed us to point out areas for future research.
The final study consists of an exploratory online survey with six questions where our responses were collected between May 2024 and January 2025.



\paratitle{Study domain}
We investigate how the industry of Canada approaches the adoption of novel privacy technologies and what impacts this adoption with consideration as to the potential relationship of policies such as PIPEDA~\cite{pipeda2000} and Bill C-27 (and the AIDA act)~\cite{billc27}. 
Our survey questions are designed with a focus on (i) understanding the existing data handling practices and procedures of companies in Canada, (ii) exploring the impact of regulatory frameworks on privacy practices and technological advancements, and (iii) identifying gaps and challenges in current privacy policies and technologies.

 We focus on Canada as it has had a version of a national privacy law since the 1980s when the \textit{Privacy Act}~\cite{theprivacyact} was first formulated, and has since seen revisions with PIPEDA in 2000, and a recently proposed update to PIPEDA that was a part of Bill C-27. 
 Based on the privacy law evolution that exists in Canada, we hypothesize that Canada's industries have some basic ways of handling sensitive customer or client data as part of their normal processes. We further hypothesized that the emergence of new regulations may result in companies  having concerns that their current, previously compliant, data handling practices may be insufficient and are actively seeking resources and strategies that could address their concerns.

\paratitle{Question design and study procedure}
Participants entering our study were informed that participation was voluntary and that they could withdraw at any time before they submitted the study by pre-emptively exiting the survey. They were then prompted to indicate via radio button response whether they agreed to participate or did not agree to participate in the study. If they agreed, the study proceeded to our first question and continued until all questions were answered or until the participant exited the study. The full set of questions is included as Appendix~\ref{app:survQuestions}.

We developed our short series of questions to target the four research questions (recall from Section~\ref{sect:intro}). In the following, we refer to survey questions as SQs. 
The questions in our survey build in terms of their specificity as the questions progressed. 
 Since more details relating to the goals of our study were included as the questions progressed, the order was not randomized, and each participant received the exact same set of prompts. 
 
For instance, the first question in our survey serves as a general baseline and warm-up question for our participants and asks about the current data handling practices that the respondent's organization may use for customer or client day. 
The survey then builds up, with SQ2 inviting explanations of what resources or plans the company considers when new privacy regulations emerge, SQ3 asks about compliance, SQ4 queries the decision-making process for the adoption of PETs, SQ5 requests examples of privacy technologies that they perceive as being widely adopted, and finally SQ6 inquires as to what factors ease the use of privacy technologies for their industry. 
Overall, completing all six questions takes participants approximately 15 minutes.

In developing our questions, we ensured that our design investigates what technologies the industry perceives as privacy-preserving, even if they don't match what research experts in privacy would classify as PETs. We have made an intentional effort in our survey not to predefine these notions, thereby reducing bias.
Finally, before releasing our study we ran a small pilot on ``similar'' relevant folks prior to releasing the survey as part of our development process. No issues emerged with the study during this pilot.


\paratitle{Participant recruitment}
We primarily recruited our participants through a non-university mailing list (a not-for-profit institute in the researcher's country) as well as through the researcher's own networks of industry folks. This email list includes 170 contacts of which 143 engaged with the email in some fashion. Of those that engaged with the email approximately 55\% have under 200 employees and the remainder have over 200 employees. Of those contacted, the majority are in the information technology sector followed by the health sector and then the financial sector.
Two participants were recruited through the sharing of a QR code at an event about new regulations in Canada where the majority of participants were from companies in the technology or financial sector. The rest of the potential participants, recruited for saturation testing, received an email from the researchers. We contacted seven potential participants for our saturation check of which five were non-senior in their careers while two were senior decision makers. Among the seven contacted, three were from the financial sector and three from the technology sector. Finally, five of the total potential participants contacted in the check stage have privacy technology experience and in terms of customer size three were from organizations with under 200 employees while four were from organizations with over 200 employees. 

The email, whether from the researchers or sent via the non profit organization, described the research and provided a link to our short survey. Following the link would bring potential participants to our consent information, and then, if they so choose, they could proceed to answer our survey questions or exit the link. 
Participants were informed clearly that there was no expectation or requirement to participate and that they could quit the study at any time during the survey. To ensure there was no power differential between the researchers and those recruited, the survey was completely anonymous such that we are unable to determine who did or did not complete the study. 

While we did not collect personal demographic data, we only recruited those who did satisfy our requirements. Participants were to be between the ages of 18 and 65, working in industry in Canada, and English speaking. 
We limit participants to adults working in Canada as the focus of our investigation is the Canadian industry's approach to compliance with privacy regulations. We restrict to English speaking as that is the language in which the study will be administered. 

Our initial recruitment via mailing lists resulted in responses from 16 participants. We then recruited an additional six participants using our personal industry networks. These additional six were selected from different industries and in different roles to test our proximity to reaching thematic saturation~\cite{glaser2017discovery}, at which point after the six participants' responses were included, we found that saturation was reached and we stopped recruitment at our total of $N=22$ participants. We assigned each participant in our final set a random identifier between one and 60, and referred to them throughout as P\#, where \# is their identifier. 

We note that being a privacy expert is not a requirement for our participants as we are trying to understand industry’s perceptions and beliefs about PETs and not to assess their expertise. Rather than evaluating whether or not ``industry'' has the correct views and whether they are ``qualified'' to speak on these things we focused on hearing openly what industry members actually think about and respond. 
Quality of responses was very rich with participants providing several sentences to paragraphs per question as their responses.


\paratitle{Research team}
Our team includes industry experience as a cryptography consultant as well as expertise and experience in qualitative coding and security and privacy technical expertise. Collectively our team holds expertise relevant to creating and assessing PETS, industry and PETs, and qualitative research.

\paratitle{Qualitative analysis} 
We analyzed participant responses using an inductive approach to allow themes to emerge~\cite{saldana2021coding}. Responses were organized first by survey questions. Then, the team of researchers analyzed all participant's responses to that question and collaboratively formed clusters, which became our sub-themes. Then, we analyzed these sub-themes together with the survey question themes viewed collectively with respect to the research questions that the survey questions were targeted at. This second phase of analysis produced the final themes that are stated in our results and presented as an overview in Table~\ref{tbl:themes}.


\paratitle{Ethics} The full study, including recruitment emails and documents, the survey questions, and our data handling, were reviewed by our institutions' equivalents of IRB. We ensure anonymity by not collecting personal identifiers. All questions were optional and participants could choose to not answer any question. The participants could withdraw by exiting the survey early, without completing it. All of this information was included in the consent form the participants were provided with before they could proceed to complete the survey. As our target participants were industry folks who would be discussing their business practices and norms we wanted to ensure that we could keep their participation private and not have any undue pressure on them. Therefore, we kept our survey short (six questions) and did not provide remuneration that could otherwise connect our participants to our survey. Thus, participants in our study's only benefit is to contribute to science that works towards a better understanding of what fosters or impedes the adoption of privacy technologies by industry.

\paratitle{Limitations}
Our study focuses on industry within Canada and thus has a Canada-based population, which correspondingly, is a WEIRD population~\cite{schulz2018origins}. While this is a limitation of our study, it is a deliberate one, to allow us to focus on an understudied region with established and emerging privacy norms and regulations. Our responses may have been impacted by the timing of our data collection as most of our responses came in while Bill C-27, which was a privacy law, was under consideration by the Canadian government. We cannot predict how the potential changes that would have resulted had C-27 passed could have impacted our participants or how it may have made regulations more prevalent in participants' minds. Correspondingly, we cannot know how it could have impacted the participants' perceptions on what responses would be more socially desirable, a factor that persists in human-response based empirical studies~\cite{redmiles2018asking}.

\begin{table*}
\centering
\scalebox{0.7}
{
\begin{tabular}{@{}ccccc@{}}
\toprule
\multicolumn{5}{l}{\textbf{RQ1 Themes and Sub-Themes}} \\ 
\midrule
\multicolumn{5}{l}{\textbf{Theme: Evaluate the relationship to their organization}} \\
 Compliance & Relevance & Cost evaluation & Market loss & Risk evaluation \\
\multicolumn{5}{l}{\textbf{Theme: Monitoring for new guidance and reports}} \\
 & (Official) Summaries & Review legislation & News letters &  \\
\multicolumn{5}{l}{\textbf{Theme: Identify necessary updates by departmental group}} \\
 & IT systems & Employee training & Control systems/processes &  \\
\multicolumn{5}{l}{\textbf{Theme: Consult with appropriate experts}} \\
Legal team & Outsource/external counsel & Privacy team & Risk management team & Certification \\
\midrule
\multicolumn{5}{l}{\textbf{RQ2 Themes and Sub-Themes}} \\ 
\midrule
\multicolumn{5}{l}{\textbf{Theme: PETs functionalities’ compatibility impacts adoption}} \\
 Adoptability & PETs design features & Amount of data & Learnability & Usability \\
\multicolumn{5}{l}{\textbf{Theme: The industry’s socio-economic system plays an important role in PETs adoption}} \\
 & Business logic & Regulations & Validation & Audits  \\
\multicolumn{5}{l}{\textbf{Theme: Cost-benefit evaluation forms the process}} \\
Roadmap  & Risk management & Resources & Education & Verification \\
 \midrule
 \multicolumn{5}{l}{\textbf{RQ3 Themes and Sub-Themes}} \\ 
  \midrule
\multicolumn{5}{l}{\textbf{Theme: Decision-making process for PETs adoption varies across industries and organizations}} \\
  \hspace{0.3cm}Company size & Dedicated teams & New Initiatives Review & Regular Monitoring &  {Consultation}  \\
 Data risks & Roadmap & Regulations & Reputation &  Best practices \\
  Clients & Existing services \& vendors & Business leaders & Competitors &  Legal consultants \\
  IT & Development team & Cyber-security & Compliance &  Decision makers \\
 \multicolumn{5}{l}{\textbf{Theme: There are varied practices, not necessarily PETs, for collecting, storing, analyzing, and utilizing sensitive data}} \\
      Outsourcing & Approval processes & Data retention policies & Performance monitoring &   Anonymization \\
  \multicolumn{5}{l}{\textbf{Theme: There are some PETs that are perceived as widely adopted}} \\
 & Encryption & Access control & Password management &    \\
 \midrule 
 \multicolumn{5}{l}{\textbf{RQ4 Themes and Sub-Themes}} \\ 
\midrule
\multicolumn{5}{l}{\textbf{Theme: Organizations report various personnel for reviewing privacy compliance}} \\
     \hspace{0.3cm}Legal advisors & Internal officers & Designated team &  Government advisors &  Point of Contact for Service \\
  \multicolumn{5}{l}{\textbf{Theme: Compliance validation includes changes across a range of internal processes}} \\
    & Training & Policy Updates & Transparency &  \\
   \multicolumn{5}{l}{\textbf{Theme: Process desc.s include a reliance on external organizations rather than specific PETS for privacy compliance}} \\
     & AWS & Microsoft (365) & Norton &  Credit card company\\
\bottomrule
\end{tabular}
}
\caption{This table provides an overview of our themes and sub-themes that emerged during our analysis.}\label{tbl:themes}
\end{table*}

\section{Results on regulation's impact}\label{sect:RQ1} 
Our first research question, RQ1, focuses on understanding how regulation influences industry adoption of privacy technologies. In this section we present our results organized by the themes that emerged from the relevant survey questions. These themes emerge from SQ2 which prompted respondents to share what resources or plans their organization considers when faced with new regulations. 

\paratitle{Evaluate the relationship to their organization}
In terms of the relationship regulation has to organizations, participants reported on the importance of compliance in their companies, their views on whether they were beholden to such regulations, and the risk their organizations faced in terms of preserving business processes in the face of changing regulations. 

Participants in our study emphasized that any applicable regulations require their compliance, and that ensuring they fulfill the requirements of such regulations is something they ensure. The importance of such compliance is emphasized by P23 who stated:
\begin{quote}
    \textit{``We fully understand the importance of these regulations from both a user and business perspective and are committed to aligning our operations with these standards as we continue to grow.'' (P23)}
\end{quote}

Others even  reported learning from regulations, and that they inspired their organization to minimize what they collect such as was the case for P22 in regards to GDPR who stated \textit{``that if we don’t collect personally sensitive data, we avoid a lot of risk to our business.'' (P22)}. 

While participants emphasized they were aware of the importance of compliance, participants also mentioned that they collect data that may be exempt. For instance, \textit{``
We are not directly affected by those regulations, however our clients are affected by those regulation changes''}(P54). Such exemption views include that the data, despite being sensitive, is \textit{``not under the purview of these frameworks''} (P43) and that the regulations \textit{``don't apply to us until the [provincial government] adopts them and formally requests we adhere to them'' }(P45).  

The final relationship vector discussed by our participants was the potential for it to hinder their business processes. Some expressed concern that they would no longer be able to meet customer demands and that \textit{``...we start to get handcuffed operationally if we have to restrict access''} (P40). 

Collectively, participants do report that they make efforts to comply with relevant regulations. However, they also report that they do not collect much sensitive information or that the sensitive information they do collect is not necessarily regulated by such laws.

\begin{tcolorbox}[colback=gray!30,
                  colframe=black,
                  width=\columnwidth,
                  arc=3mm, auto outer arc,
                 ]
    The conflicting views on what is sensitive data and the applicability of regulations, imply a need for clear guidance on what is and is not regulated and what does and does not satisfy compliance to ensure organizations are not inadvertently non-compliant. 
\end{tcolorbox}


\paratitle{Monitoring for new guidance}
Participants describing their process mentioned a selection of sources as being where they turn to for guidance. 
When going to official sources such as legislation, participants report it taking extensive reviewing of both the legislation and government summaries. One example provided by a participant about GDPR discussed that reviewing the regulation \textit{``...required a lot of work to click through each of the recommendations, but...I chose to ensure that we were at least current with protocols that were ahead of our legal requirements''} (P22)

Two strategies were reported for finding easier paths to understanding the regulation instead of spending extensive time on government sources. P21 reports that they first, \textit{``review the summaries of the legislation that are put out and if needed review the legislation''} (P21). Other participants have made an effort to acquire additional sources, for instance signing up \textit{``to a few privacy newsletters that allow bite sized consumption of legislative updates across our markets, and we choose to set up notifications for those that appear of interest''}.


\begin{tcolorbox}[colback=gray!30,
                  colframe=black,
                  width=\columnwidth,
                  arc=3mm, auto outer arc,
                 ]
Processes do not necessarily rely only on accessing only official government web pages, but also less official, though potentially more intuitive, sources.  Our participants refer to government web pages and privacy newsletters as resources they employ to learn how their market is impacted by regulations. 
\end{tcolorbox}

\paratitle{Identify necessary updates by departmental group}
Normal processes, according to our participants, include consideration for the impact structured by the department group. 
The regulations can cause changes that propagate throughout the entire organization as \textit{``New privacy regulation may result in our internal policy documents and all applications and business lines must follow the policies''}(P51).
Determining which departments need what updates may be done through a \textit{``...mapping of the requirements against our frameworks and programs, identification of gaps and plans to make the required updates to support ongoing compliance''} (P4). Ensuring compliance to new regulations includes updates to technical systems, as highlighted by P24 \textit{``...we assess and update our IT systems and processes to meet the new regulatory requirements, including enhanced data protection features or new consent management tools''}.
However, the impact on business processes also extends to other internal facing and customer facing departments. Updates are undertaken to ``website, marketing, lab services, and software'' (P21) as well as to ``employee training programs'' (P24), and collection practices using ``cookies and advertising platforms'' (P20).

\begin{tcolorbox}[colback=gray!30,
                  colframe=black,
                  width=\columnwidth,
                  arc=3mm, auto outer arc,
                 ]
Updates made in organizations to adapt to new regulations span departments, including less visible ones such as employee training, in addition to more directly impacted ones like IT, legal, and marketing. 
\end{tcolorbox}

\paratitle{Consult with appropriate experts} 
The personnel that are consulted with in regards to how to adapt to new regulations has variations. For some organizations, consultation consists of reviews by their internal \textit{``risk management team''} (P36). Some processes explicitly refer to the vagueness as a factor for why \textit{``The regulations are reviewed by lawyers with a background in privacy''} (P55). Those without such teams may choose to reach outside their organization and \textit{leverage an audit/conformance/assurance partner'' (P16)}. Hiring out this task can include having a consulting company that 
\begin{quote}
    \textit{``translates the new regulations and hands them down to us who then explains it to our staff in a more comprehensive manner. The information from the government is FAR to advance for the people who are actually required to do the work''} (P52).
\end{quote}
Organizations that are very new, such as startups, may not have internal or external entities they already use. Instead, they may plan to pursue it through formal processes in the future such as to \textit{``...pursue these certifications in the next 6-12 months''} (P23). 



\begin{tcolorbox}[colback=gray!30,
                  colframe=black,
                  width=\columnwidth,
                  arc=3mm, auto outer arc,
                 ]
                 
 Expert consultation can be part of the decision making process. However, the size of the company and how long the company has existed impacts whether it can effectively employ consultation with external experts, outside consultants, or require some future process that they have not yet established.
\end{tcolorbox}

 


\section{Results on factors for PETs adoption}\label{sect:RQ2}
Our second research question, RQ2, searches for factors that the industry perceives as important in determining whether they adopt privacy technologies. The following three themes emerged from the analysis of SQ6, which explicitly asked participants what factors influenced the adoption of privacy technologies in their industry.  


\paratitle{PETs functionalities' compatibility impacts adoption} 
Our participants highlight both features of PETs that they would view as beneficial in terms of functionality as ones that they view as easing the feasibility of adoption. 
One aspect that participants report impacting adoption is the effort required to integrate it into their systems. For instance, whether the system is low-impact \textit{``in terms of time and energy''} (P22) or just generally has \textit{``ease of use''} (P55 and P56). Similarly, it should be easy to determine what it can work with as otherwise \textit{``it takes a lot of time to discover whether or not different tools play nice together, especially across platforms''} (P22). 
The functions need to be compatible with the business goals and support the organizations' \textit{``ability to continue to carry out our mandate'' }(P44) which requires there to be \text{``accuracy''} (P55) as well as \textit{``robustness''} (P22), and be both \textit{``scalable''} (P23) and \textit{``affordable''} (P23, P55, P56). 

\begin{tcolorbox}[colback=gray!30,
                  colframe=black,
                  width=\columnwidth,
                  arc=3mm, auto outer arc,
                 ]
                 
 Ease of integration in terms of time, money, and functionality compatibility with the organization remain factors for technology adoption, including for PETs.  
\end{tcolorbox}


\paratitle{The industry's socio-economic system plays an important role in PETs adoption} 
In terms of PETs adoption, regulatory requirements can encourage the deployment of PETs (P4, P22, P23, P32, P36, P37, P44). However, this encouragement is not unilateral. Participants expressed that for regulations to encourage PETs adoption, the PETs \textit{``technology must comply with relevant privacy regulations''} (P24). While regulation can encourage, participants report several other aspects of the industry's socio-economic system that impact the adoption of PETs.
The industry's socio-economic system encompasses its regulatory compliance, market dynamics, interactions with competitors, partners, and emerging innovations that shape its overall position.

Market dynamics, including elements such as client demands, domain expectations, competitors practices, and domain perceived best practices emerged as factors our respondents are influenced by (P21 , P23 , P43, P46). Value was placed on standardization practices, not just by standardization organizations like NIST (P51), but also emergent standards in the form of norms for that field of industry. Emergent standards become not just as safety expectations but also factors that encourage clients to use that business as stated by P23: 
    \begin{quote}
        \textit{``if a particular privacy measure is becoming standard in the industry or is being sought after by our clients, it reinforces the need for us to adopt it. Not only do these technologies help safeguard us from potential security breaches, but they also serve as a key selling point for our clients''} (P23). 
    \end{quote}

    While the interplay with market dynamics was stated as a motivator for adopting PETs to boost business value and reputation (P36, P44), it was also reported as an impedance for the deployment of less known technologies. For instance, concerns were with how to convey the value in the adoption of PETs deployed by larger competing companies:
    \begin{quote}
        \textit{ ``It is important that we educate our customers on why we are using different technologies that are not adopted fully in big corporations (especially internationally)''} (P21). 
    \end{quote}

Further relating to organizations needing to account for what large industry leaders are doing, some report looking to those same large organizations as guides on standards and practices. The views include that since such organizations face bigger consequences for failure they need to pick good practices, or said another way \textit{``large companies are always rightly concerned about data privacy''} (P21).
In addition to looking to other organizations to lead, consultants are also employed when considering what best practices are, something that can aid in efficiency for some organizations:    
    \begin{quote}
        \textit{``I rely quite heavily on the [consulting] company I hired. I just simply don't have the time it requires or the resources to successfully stay on top of all the new legislation ... and cyber-security [defences]''} (P52).
    \end{quote}

\begin{tcolorbox}[colback=gray!30,
                  colframe=black,
                  width=\columnwidth,
                  arc=3mm, auto outer arc,
                 ]
                 
The socio-economic system in which companies exist influences their willingness and ability to adopt PETs. Challenges include what competitors are doing, being able to justify using novel technology that well known large companies are not using, and the time to keep up with new changes and threats. 
\end{tcolorbox}

\paratitle{Cost-benefit evaluation forms the process}
In the previous theme we discussed how there were several risks that participants expressed concern with, such as the perception of their clients if they used unusual technology. 
Risks and challenges exist. Correspondingly, it is also the case that a technology's ability to effectively address a data protection need and mitigate the risk of harm and reputation (P16, P23, P24, P36, P43) is necessary for its deployment. While necessary, there are still additional factors that go into the decision process beyond that risk mitigation, as the mitigation is not sufficient on its own. Rather, in addition to successful mitigation, the organization needs to \textit{``... assess whether the technology makes sense from a product and business perspective''} (P23). This assessment can be even more formal when determining whether the overall cost-benefit trade off is worth it, such as using a  \textit{`` weighted approach to determine which technologies will deliver the greater net benefit for our user base'' }(P32).

Among the costs considered by the participants, there are budgetary concerns such as \textit{``the timing, budget, and resources required to implement it effectively''} (P23). In addition to these more conventionally budgetary concerns, there are also concerns for verification needs for organizations that
\textit{``conduct regular audits to ensure infrastructure and servers are secure''} (P43). 
Other costs include efforts to ensure buy-in from both employees and decision makers:
\begin{quote}
    \textit{``Securing stakeholder buy-in helps ensure the technology receives the necessary resources and backing within the organization...Providing training and raising awareness among employees about the new privacy technologies [also] ensures they are used correctly and helps maintain overall data security''} (P24).
\end{quote}

\begin{tcolorbox}[colback=gray!30,
                  colframe=black,
                  width=\columnwidth,
                  arc=3mm, auto outer arc,
                 ]
                 
While addressing a data protection need is essential in considering PETs for deployment, there are many other factors that affect their deployment.
\end{tcolorbox}


\begin{figure} 
\centering
\resizebox{0.8\textwidth}{!}{
\begin{circuitikz} 
\tikzstyle{every node}=[font=\large]
\draw  (4,15.75) rectangle (17.25,10.9);
\draw  (4,10.9) rectangle (17.25,7.6);
\draw  (4,7.6) rectangle (17.25,5.1); 
\draw  (4,5.1) rectangle (17.25,3.5);
\draw  (4,3.5) rectangle (17.25,1.9);
\draw [ rounded corners = 24.0] (4.25,15) rectangle (13.25,11);
\draw [ rounded corners = 24.0] (14,14.25) rectangle (17,11.25);
\draw [->, >=Stealth] (13.25,13.25) -- (14,13.25);
\node [font=\Large, align=center] at (11.25,15.4) {\textbf{(i) Need for PETs}};
\node  at (9,14.75) {\textbf{Monitoring}};
\node at (15.5,13.75) {\textbf{Validation}};
\node [align=center] at (15.5,12.5) {- Compliance \\ - Client agreement \\ - Market Position\\ - Addressing risk};
\draw [ rounded corners = 12.0, align=center] (4.5,4.4) rectangle  node { Initial approval} (7.5,3.6);
\draw [ rounded corners = 12.0, align=center] (9.6,4.4) rectangle  node { Pilot test} (12.8,3.6); 
\draw [rounded corners = 12.0, align=center] (13.95,4.4) rectangle  node {Final approval} (17,3.6);
\draw [->, >=Stealth] (7.5,4) -- (9.6,4);
\draw [->, >=Stealth] (12.8,4) -- (14,4);
\node [font=\Large, align=center] at (11.25,4.75) {\textbf{(iv) Approval}};
\draw [ rounded corners = 12.0, align=center] (5.5,2.8) rectangle  node {Implementation plan} (9.5,2);
\draw [ rounded corners = 12.0, align=center] (12.9,2.8) rectangle  node {Performance monitoring} (17,2);
\draw [->, >=Stealth] (9.5,2.4) -- (12.9,2.4);
\node [font=\Large, align=center] at (11.25,3.1) {\textbf{(v) Post Approval}};
\node [font=\Large] at (11.25,7.25) {\textbf{(iii) Evaluation} };
\draw [ rounded corners = 7.5] (4.25,6.9) rectangle  node {\textit{Internal:} policies/standards, function, work flow compatibility, UX, business goals} (17,6.4);
\draw [ rounded corners = 7.5] (4.25,6.3) rectangle  node {  \textit{Safeguarding and defense:} client perception, rep., risk mitigation, achievable sec.} (17,5.8);
\draw [ rounded corners = 7.5] (4.25,5.7) rectangle  node { \textit{Costs: }time, budget, resources, training, ease of deployment...} (17,5.2);
\node [font=\Large, align=center] at (11.25,10.5) {\textbf{(ii) Exploring Solutions}};
\draw [ rounded corners = 24.0] (4.25,10.2) rectangle (17,8.3); 
\draw [ rounded corners = 7.5] (4.25,8.2) rectangle  node { 
If not exist, possibly need to make it in-house, internal solution} (17,7.7);
\draw [ rounded corners = 7.5] (6.2,10.1) rectangle  node { \textit{Who is using it:} business leaders, competitors, domain...} (16,9.6);
\draw [align=center, rounded corners = 7.5] (6.25,9.5) rectangle  node { \textit{Who makes it:} reputation, already a vendor, ...} (16,9);
\draw [ rounded corners = 7.5] (6.25,8.9) rectangle  node {\textit{What is recommended:} client demand, recommendations...} (16,8.4);
\node at (5.3,9.75) {If exists: };
\draw [align=center, rounded corners = 15.0] (4.4,14.5) rectangle  node {\textit{Internal Personnel: }Cyber security team, subject matter \\experts, IT administrator,development team} (13.1,13.5);
\draw [align=center, rounded corners = 15.0] (4.4,13.4) rectangle  node {
\textit{Required changes:} compliance with regulation, ~~~~~\\requirements for organizations in a jurisdiction} (13.1,12.4);
\draw [align=center, line width=0.5pt , rounded corners = 15] (4.5,12.3) rectangle  node {
\textit{Organization needs:} vulnerability, data risk, regulation \\ reqs, client demands, competitors, security as a value} (13.1,11.3);
\end{circuitikz}
}
\caption{Decision-making process for PETs adoption as captured from the respondents. The process flows from top to bottom for the large rectangular steps and proceeds through the different components within each larger overall step. }\label{fig:PETS_Process}
\Description[Flows from step i to iv with sub-steps described]{Shows decision-making process for PETs adoption as captured from the respondents. The process flows from top to bottom for the large rectangular steps and proceeds through the different components within each larger overall step.}
\end{figure}
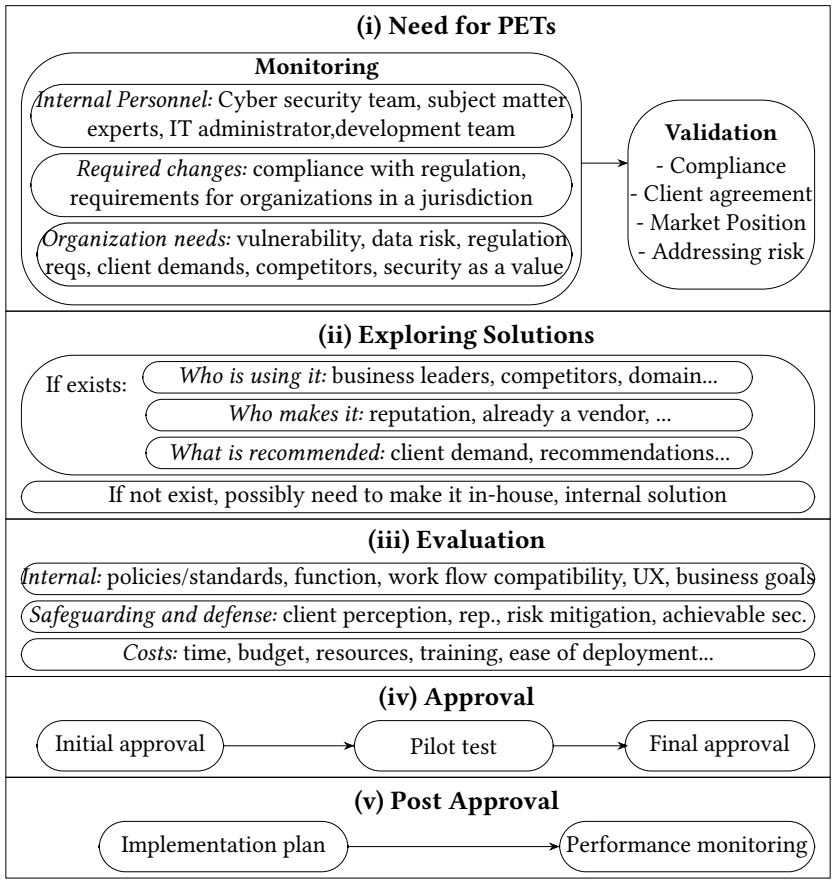

\section{Results on common practices regarding PETs}\label{sect:RQ3}

Our third research question, RQ3, studies the common practices for PETs by industry. We synthesize the results for this question which consist of the emergent themes from SQ4 and SQ5. SQ4 and SQ5, respectively, asked participants about their decision-making process for adopting or developing PETs and an example of what participants thought is a (relatively) widely adopted PET in their domain of industry. We formed a collective diagram of the PETs adoption steps by incorporating the descriptions from each of our participants, which we include as Figure~\ref{fig:PETS_Process}. We will refer to Figure~\ref{fig:PETS_Process} throughout this section when it relates to the theme under discussion.   


\paratitle{Decision-making process for PETs adoption varies across industries and organizations} There is a wide range of responses in our study, spanning from \textit{``our decision-making process is a collaborative effort between our CTO, development team, and leadership''} (P23), to \textit{``our decision-making process for adopting or developing privacy technologies involves several key steps'' }(P24). Each company undergoes a sub-diagram process based on their size and structure. For example, some responses suggest that PETs adoption decision-making in smaller companies may require fewer steps, and involve the management earlier in the process: 
\begin{quote}
    \textit{``Due to our smaller size as a company, [PETs adoption] decisions are handled among the management team directly with no real formal process. It is left up to the individual teams to think of privacy'' }(P40).
\end{quote}

In addition to potentially including management earlier, other small organizations attributed their size as a contributor to how quickly they are able to implement changes: 
\begin{quote}
        \textit{``We have an IT Administrator that reviews industry best practices regularly and recommends changes to the CTO. These changes can be reviewed and implemented quickly because we are a small organization''} (P21). 
\end{quote}

The decision making process, while varied, has five steps we have identified and summarized in Figure~\ref{fig:PETS_Process}. To understand the factors that go into each of the five steps, we present the following detailing of them with examples. 
\paragraph{(i) Need for PETs.} Regardless of speed, what the full process is for determining whether to adopt PETs has variation, with the first step having some form of monitoring the status quo and organizational needs (see the first block in Figure~\ref{fig:PETS_Process}). 
The need may be identified during a regular procedure such as \textit{``following consultation with industry and internal [decision makers]''} (P44) or alternatively because changes are \textit{``legally necessary''} (P37).
Those involved in the instigation process can be IT professionals (P21, P53, P54), development team, cyber-security (P44, P53), or subject matter experts (P36). Some of the driving factors in this step are internal, such as addressing an identified vulnerability or a data risk (P16, P24, P32, P36), acting to support road-map (P16), and even just to be as secure as possible (P24). Other motivators are more external such as regulations (P24, P37, P38, P44, P53) and standards (P45), client demand/benefit and reputation (P23, P32, P38, P43), consultation and best practices (P38, P44, P46, P53, P54).
Finally, even before proceeding further in the process, the technology's ability to assist with regulation compliance (P24, P38, P44), mitigating risks (P4), boosting the industry's market position (P23), or gaining client's approval (P43) is used to validate the need before moving on to explore solutions viability. 

\paragraph{(ii) Exploring Solutions.} The exploratory task of searching for available mitigation solutions to the identified need can be influenced by factors inside and outside of the organization. 
Outside factors include client demand (P23, P43), and recommendations by existing service providers (P46) while internal factors come through various personnel such as \textit{``internal technology and business leaders''} (P16). 
Some personnel, such as in the case of organizations with dedicated teams, make direct suggestions where 
\textit{``recommendations come forward from our cyber-security team'' (P44).}
Other experts may also suggest solutions, even if they are not part of a dedicated team where \textit{``subject matter experts suggest mitigation plans''} (P36).

Other factors that determine which solutions are explored include market dynamic elements such as business domain practices (P46), competitors (P23), industry best practices (P38, P44), and consultation with legal experts (P44).
In some cases, vendor evaluation might even precede the solution search part of the process \textit{``if considering third-party solutions, we assess potential vendors based on their technology, reputation, compliance with industry standards, and support offerings''} (P24).

\paragraph{(iii) Evaluation.} After the initial solution exploration, and possible vendor evaluation, the next step is to evaluate the potential solutions. For this stage, we refer back to the results in Section~\ref{sect:RQ2} where the same evaluation factors were synthesized by theme as we found in reports on participants' decision-making process. In short, the evaluation process includes cost-benefit analysis, consideration for internal needs, and the compatibility of the technology with the business goals of the organization.

\paragraph{(iv) Approval  and (v) Post Approval} The process reaches a conclusion when a solution gains preliminary agreement from various stakeholders including IT, security, compliance, as well as user representatives (P21, P36, P53). This preliminary agreement is termed the initial approval stage in our diagram.  After the initial approval is achieved, the next requirement is to demonstrate success through testing (P24, P53). If the testing is successful, then the proposed mitigation solution secures final approval from executives and decision-makers (P21, P24, P44). Once the decision is finalized with all stages of approval, an implementation plan is devised (P24, P36, P55), and the solution's performance is closely monitored during full deployment (P24). Afterward, the solution may receive continued monitoring and could loop back to the beginning of the process shown in Figure~\ref{fig:PETS_Process}, where you once again determine if there is a need for new PETs.


\begin{tcolorbox}[colback=gray!30,
                  colframe=black,
                  width=\columnwidth,
                  arc=3mm, auto outer arc,
                 ]
                 
The decision-making process for whether to consider and whether to deploy PETs has a variance that can be attributed to both the industry domain and the size of the organization. This can serve as a reminder that there will be no one-size-fits-all solution for easing the adoptability of PETs for industry actors.
\end{tcolorbox}


\paratitle{There are varied practices, not necessarily PETs, for collecting, storing, analyzing, and utilizing sensitive data} 
When prompted to describe a privacy-preserving technology method that they thought of as widely adopted, some participants described things which researchers and practitioners in the community would not necessarily think of as PETs. 

One such non-PET mentioned is testing, where the testing is \textit{``conducted with a limited user group to evaluate effectiveness, usability, and integration''} (P24). While we can agree that testing is an important part of any system change, it is hard to attribute the testing itself with a privacy-preserving function. 
Approval processes, where material and information \textit{``must be approved''} (P21) before being shared or disclosed are reminiscent of access control processes, and so move us closer to a PET. 
A not necessarily technological version of access control is also mentioned as a way to protect data, where the organization's solution is \textit{``Have Structural restrictions to limit peoples ability to access sensitive information internally''} (P22). 
What we found is that while some participants may not have mentioned specific PETs or strictly defined PETs, they do report various types of controls that are used; including, for example, that
\textit{``data retention policy is another controller that is commonly used to ensure compliance with privacy mandates and regulations''} (P36).

Responses also mention \emph{de-identification and anonymization} (P4, P21, P51, P56) including \textit{``anonymization network which works to advance certainty in the industry around how to do this well''} (P4) and \textit{``data that is published in paper in anonymized''} (P21). However, it is well known that simply anonymization does not guarantee privacy, but this is still a common practice in the industry.
  

In addition to the non-PETs mentioned, participants also mention a solution that is not itself a PET but is a way to procure PETs. That is, one of the solutions put forth by our participants is to outsource to someone else, such as by using \textit{``Secure infrastructure as offered by the common cloud vendors''} (P37), \textit{``antivirus software''} (P22), \textit{``secure file transfer''} (P36, P53), and \textit{``third-party payment processing''} (P40). 



\begin{tcolorbox}[colback=gray!30,
                  colframe=black,
                  width=\columnwidth,
                  arc=3mm, auto outer arc,
                 ]                 
Industry's approach to privacy-enhancement is not necessarily what experts and researchers perceive that it should be. However, there are many strategies employed by industry that they view as being for privacy protections and understanding those processes may guide our understanding of what type of PETs industry needs. 
\end{tcolorbox}




\paratitle{There are some PETs that are perceived as widely adopted}
While non-PETs were also mentioned in regards to privacy preservation, more conventional privacy-enhancing techniques were also mentioned by participants. Participants report standardized techniques such as encryption as well as techniques such as access control and password management.

Encryption techniques reported support both data storage and transfer. Encryption has some very specific examples provided by participants such as using \textit{``as 256-bit encryption, which protects data both at rest and during transfer''} (P24) and \textit{``SSL/TLS''} (P46).
Specific technologies that employ encryption and facilitate the organization's business processes were highlighted as well.
\begin{quote}
    \textit{``a huge privacy preserving technology our industry uses would be DocuSign, all documents sent through DocuSign are encrypted''} (P53).
\end{quote}

Descriptions in the responses include high-level interpretations about access control such as \textit{``Have Structural restrictions to limit peoples ability to access sensitive information internally''} (P22).
However, there are no mentions of specific tools that support access control and that granularity of control. 
There is one detailed example given regarding access control. P32 gives a detailed example, 
\begin{quote}
    \textit{``Intentional regionalization of data. In order to ensure customers maintain the privacy rights afforded to them by where they reside, we offer hosting of our services in multiple geographies, with no data-transfer methods built to move data across these boundaries. We then build our product and privacy practices around the strongest of privacy regulations to allow ourselves a buffer as other jurisdictions adopt parts of these strong privacy regulations.''} (P32).
\end{quote}
While the description is quite detailed, we do note that it is unclear if the company is using particular in-house database tools or using vendors to support their desired access control policies. We also cannot determine how these practices can meet privacy regulation requirements.

Finally, participants in our study also mention a selection of  specific PETs, authentication assistants, password managers, and password assistants. 
In terms of password management, some mention the practice is company wide, with there being \textit{`` enterprise wide password keeper''} (P22) as part of their standard practices.
Other access based solutions reported include  \textit{``VPN secure sign in.''} (P44) and third party access management systems \textit{``for role based account access''} (P46).


\begin{tcolorbox}[colback=gray!30,
                  colframe=black,
                  width=\columnwidth,
                  arc=3mm, auto outer arc,
                 ]
Encryption, access control, and password related solutions are among the PETs that participants mention when asked for ones that are widely deployed. 
\end{tcolorbox}


\section{Results on compliance verification}\label{sect:RQ4}
The fourth research question, RQ4, is to understand how industry determines whether they are compliant with relevant regulations and policies. The results for this research question are organized by themes that emerged from the analysis of SQ3, with additional discussion of relevant themes from SQ5 and SQ6. 


The responses from SQ3 on the processes industry has for compliance has two overarching types. Some responses include descriptions of very structured approaches and list concrete steps to ensure compliance with existing or emerging privacy regulations. 
In contrast to the structured approaches, there are those with \textit{``no official processes, and this is a known gap.''} (P40). Others are somewhere in between where we do not know the exact steps they undertake, but it is suggested that such steps may exist
\begin{quote}
    \textit{``We have a very organic process that involves reviewing the legislation and knowing how to navigate issues for our clients''} (P20)
\end{quote}

These approaches, whether detailed or ambiguous, can depend on the resources and priority of the organizations. For instance, as we will discuss next, the detailed steps include different personnel in the process, personnel that not all organizations will have. 


\paratitle{Organizations report various personnel for reviewing privacy compliance} 
Compliance and its verification is not an isolated task. Participants in our study report a breadth of roles that are responsible for facilitating both compliance efforts and confirmation of compliance. 
The roles may be filled by internal personnel, external personnel, or a combination of both internal and external personnel. 

The internal team can have a designated role, where \textit{``Risk management monitors the change in the landscape and will assess the impact of the regulations''} (P36). They can also be responsible for maintaining privacy and compliance more broadly or as a point of contact for regulatory authorities and individuals.
\begin{quote}
    \textit{``There is an internal compliance team that maintains the privacy program, reviews emerging regulations, and has internal audit capability to ensure the organization is ready for emerging privacy regulations''} (P32).
\end{quote}

Some internal teams collaborate with external partners and clients to review the legislation/government/sector actions (P38, P20, P43, P54, P56). 
\begin{quote}
    \textit{``Our in-house data team monitors the changes and works with our external partner to understand what is applicable to us. We allocate resources based on the priority of implementing these changes in the recommended order''} (P38).
\end{quote}

In terms of strictly external personnel, some respondents report that they have trusted legal advisors for consultation and collaboration (P16, P21, P23, P52). They state that the value of such personnel corresponds to the risk of failing to meet these requirements:
\begin{quote}
    \textit{``if we were to ever miss something from the CRA or PIPEDA we would be fined so fast. So we took the step to hire someone who will never put us in that position''} (P52). 
\end{quote}


\begin{tcolorbox}[colback=gray!30,
                  colframe=black,
                  width=\columnwidth,
                  arc=3mm, auto outer arc,
                 ]
Ensuring compliance involves a reliance on expertise from personnel, who may be on internal teams, internal personnel that consult with external experts, or completely external contractors that advise on proper compliance practices.  
\end{tcolorbox}

\paratitle{Compliance validation includes changes across a range of internal processes}
The responses include 
different classes of internal changes that could be triggered by privacy compliance.
Ensuring compliance is something that requires changes to the actual systems, not just as a way to verify compliance, but also as a way to establish new normal processes. To update these practices, some organizations ensure that \textit{``Training is provided to employees on data protection principles, privacy policies, and regulatory requirements to ensure they understand their responsibilities''} (P24). 

Further efforts include transparency through the creation and publication of \textit{`` accountability management framework''} (P4) as well as ensuring \textit{``policies are in place to ensure that data of a sensitive nature is not sent out''} (P22).
These changes are reflected both in processes and in communications to customers:
\begin{quote}
    \textit{``Our website and marketing follows best practices that our advisors and internet resources suggest. We have a standard privacy agreement that we use with customers that protects data generated''} (P21)
\end{quote}


\begin{tcolorbox}[colback=gray!30,
                  colframe=black,
                  width=\columnwidth,
                  arc=3mm, auto outer arc,
                 ]
Compliance efforts include not just changes to actual processes, but also changes to training procedures and updating communications to increase transparency to customers and clients impacted by changes. 
\end{tcolorbox}

\paratitle{Process descriptions include a reliance on external organizations rather than specific PETS for privacy compliance} 
Participants report a reliance on the personnel in organizations they outsource to for software solutions to provide any compliance assurances (P43, P23, P45, P46). Participants rely on organizations such as Microsoft Services and AWS to ensure the solutions they are providing are compliant with relevant laws.
\begin{quote}
    \textit{``Our current approach includes leveraging AWS cloud storage and their security features. We also maintain strict policies against sharing or selling personal data and ensure all survey data is aggregated and anonymized''} (P23).
\end{quote}
Responses mention the use of cloud services as the support for security and privacy, which assumes the out-sourced service, such as the cloud service provider or the credit card company, provides sufficient technology to support privacy compliance.



\begin{tcolorbox}[colback=gray!30,
                  colframe=black,
                  width=\columnwidth,
                  arc=3mm, auto outer arc,
                 ]
A reliance on outsourced solutions is one strategy for complying with regulations. However, it is unclear whether such solutions can actually ensure compliance for an organization.  
\end{tcolorbox}

\section{Discussion}\label{sect:discussion}

Our survey results create a rich picture that consists of an overview of the PETs adoption process in industry and the roles of the actors who influence it, including researchers, industry personnel, and policymakers.
To fulfill our mission of strengthening connections within the researcher-regulator-industry practitioner triangle, we share some general insights for all three parties and specific findings for each. 


\paratitle{A unique and dynamic multi-stakeholder ecosystem} Privacy and [cyber-]security go hand in hand, and both have a reliance on cryptography. Techniques such as differential privacy, homomorphic encryption, multi-party computation, and computation over encrypted data are used extensively in the state-of-the-art privacy-preserving solutions produced by research, and all have deep roots in cryptography. 

Fischer et al.~\cite{Fischer2024Challenges} provide a visualization of the cryptography ecosystem and the path of bringing cryptography from paper to product. While the technologies are intertwined, there is a significant distinction between the journey of cryptography and privacy. In the former, end users and organizations are mostly the consumers of the product, where the final arrows point to. For privacy however, all elements of the socio-economic system --- from regulations to client demand, consultants, competitors, best practices, market position, and reputation --- play an active role in various phases of the decision-making journey. We associate this difference to the nature of privacy as a social concept, the public awareness of digital technology, and its potential threats and misuses in the current era. Hence, we observe these actions unfolding simultaneously: regulators striving to translate public demand into legislation, industry working to meet market privacy needs \textit{and} ensuring compliance, and researchers developing privacy-preserving technologies that align with real-world concerns by supporting regulation and industry adoption. This interdependence can be frustrating, as each group relies on others whose work is still evolving. However, with effective communication, these efforts can align and reinforce one another. 

Compliance verification is an example of a pain point that is notably challenging at this time. To elaborate on the nature of the compliance issue, we focus on the example of anonymity and de-identification as a fundamental pillar of privacy regulations. On one hand, de-identification and anonymity is one of the main privacy topics that were brought up repeatedly in our survey responses. Many mentioned it to be one of the widely deployed privacy enhancing technologies/methods in their industry. However, as we mentioned in Section~\ref{sect:intro}, Canadian regulators are highly concerned about industry self regulating~\cite{ElEmam2024Perspectives} on the topic, since the methods industry uses for anonymity are not clear in design or implementation, and are not unified across industry. On the other hand, there are numerous research papers on how various PETs can assist with anonymity and de-identification or evaluate it. This presents an opportunity for collaboration among the three parties. Research institutes, in partnership with industry leaders, can respond to Canadian regulators' call for \textit{issuing codes of practice and certifications by non-regulatory bodies}. This can be achieved by leveraging PETs to provide a unified, systematic, and quantifiable evaluation of anonymization, such as through the use of privacy attacks like membership inference~\cite{stadler2022synthetic,Giomi2023AUnified}. 

 

\subsection{Key findings by stakeholder}
\paratitle{Key findings for research institutes}
It is advantageous that the usability of PETs is already receiving a lot of attention in the research community. 
However, as our participants pointed out, it is also crucial to consider PETs compatibility and integration with the existing platforms in industry, along with their associated cost as discussed (recall from Section~\ref{sect:RQ2}). Since regulations can accelerate the drive towards PETs adoption within industry, it would be highly beneficial to develop PETs that can be verified, audited, and demonstrably compliant with relevant regulations (recall from Section \ref{sect:RQ4}). 
The issuance of certificates by research institutes would be a collaborative effort that could help with the verification process, particularly if it aligned with regulatory requirements, which could then further support PETs adoption. The development of PETs that assist in evaluating privacy requirements will also be greatly valuable to industry actors as it will aid in facilitating regulatory enforcement with additional transparency and testability.

Understanding market dynamics in PETs adoption can help researchers navigate the challenges of bringing their work into industry. Our survey suggests that the choice of which PETs to adopt may be influenced by recommendations from existing service providers. 
While this result parallels many attributes identified in prior work on the value of networks as professional connections for knowledge or expertise sharing~\cite{ehrlich2007searching}, competitors and best practices were unique factors that were highlighted as ways new techniques come to industry's awareness and subsequent adoption. This means that not only is it the case that social influence impacts privacy behaviors of those interacting with social networks~\cite{mendel2017susceptibility}, but also that the social hierarchies of who organizations view as peers, newcomers to the field, and leaders in the field all can impact organizations in their inclination to adopt privacy technologies. Similarly, smaller companies may implement PETs more quickly due to management involvement earlier in the process, but they may struggle to convince clients to adopt solutions that lack widespread use among larger companies. Additionally, we identify the personnel departments responsible for assessing the need for PETs and exploring mitigation strategies, which are not limited to personnel on privacy focused teams. 
This suggests that an effective way of propagating PETs in industry is illustrating who among their peers uses the same or similar systems. Such a peer based grouping seems to be the closest approximation to communities of practice for our participants~\cite{wenger1999communities}. 
Therefore, a viable approach to enhance acceptance is integrating PETs with established solutions. 
As participants identify blogs and digital newsletters as a source of information, an example of computer-mediated communication that could be harnessed by the technical research community more effectively~\cite{preece2000online}, such platforms may be effective if used to communicate about usages and acceptance rates by peers.

\paratitle{Key findings for regulators} Our study suggests that industry professionals often seek clearer, more accessible communication regarding the regulations that apply to them, along with practical advice on how to achieve compliance. We note that according to a recent Canadian business survey result~\cite{phoenixspi2024survey}, only one out of four of Canadian businesses have used the information and tools provided by OPC (Office of the Privacy Commissioner of Canada) despite being aware of them. 
In our survey, respondents (recall Section~\ref{sect:RQ1}) reported government resources to be challenging to use and that they refer to privacy newsletters and other more easily understood sources when trying to decipher new requirements. 
Informed by our findings, we suggest that it would be beneficial to offer guidance that combines the authority of official sources with the actionable insights typically found in informal channels such as blogs, newsletters, or technical tutorials.
Our survey identifies the key resources that Canadian industries rely on for regulatory compliance; from domain experts to consulting firms.
Privacy regulators may find it more effective to address communication challenges by engaging with these intermediary parties, to ensure that industry receives clearer and more actionable compliance guidance. This approach could not only improve clarity but also make compliance more attainable for smaller companies that lack dedicated teams for separate tasks or face challenges in securing external counsel, which may lead them to adopt a follower strategy based on larger companies' practices.



\paratitle{Key findings for industry practitioners} 
While no outcome is more rewarding for researchers than seeing their designed PETs deployed in practice, this deployment remains a challenge. 
Our findings highlight key privacy focus areas in industry that we suggest will benefit from existing research dedicated to PETs design and development in these areas. 
While industry participants in our study employ well-established products, emerging PETs developed by researchers can offer valuable contributions; both in terms of privacy-preserving and privacy evaluation solutions. 
For example, in anonymity and de-identification context, current commercial data synthesis applications~\cite{Patki2016synthetic,gretel} either follow data sanitizing practices such as removing identifiers, or have not yet reached a level of maturity that allows them to scale effectively with the complexities and volume of industry data. However, there are several research papers advancing against the various practical problems of data synthesis~\cite{kollovieh2023selfguiding,pang2024clavaddpmm,zhang2023mixed}.

We propose that one key barrier to the adoption of emerging PETs is a lack of awareness among decision-makers responsible for implementing these technologies. 
Ensuring that personnel departments stay informed about the latest PETs advancements through education and communication can help bridge this gap. Moreover, by actively communicating their desired features to the research community, industry practitioners can also help with shaping the future of PETs, ensuring that new developments align more closely with real-world needs and constraints.

\subsection{Future Research Directions}
While this study provides valuable insights into the current process of PETs adoption in industry, it unfolds several avenues for deeper investigation and future research. 
 
\paratitle{A thorough picture of PETs used in industry, positioning state-of-the-art} This study identifies the privacy focus areas for industry practitioners and their decision making process to adopt PETs. Our participants provided several examples of widely adopted privacy enhancing technologies or methods (see Section~\ref{sect:RQ3}), but in most cases the specific technologies deployed remains unclear. For instance, multiple participants mentioned anonymization, de-identification, and encryption, yet it is uncertain whether they were referring to any of the state-of-the-art PETs currently available. 

In the context of anonymity, differential privacy has been successfully implemented by the U.S. Census Bureau and several large companies. Similarly, synthetic data generation has gained popularity in finance and health sectors. 
JP Morgan's recent 2024 report~\cite{JPMorgan2024synthetic} advocates for the use of synthetic data in finance applications, highlighting several benefits, including enhanced privacy. Additionally, a national AI institute in Canada hosted a synthetic data generation bootcamp for numerous Canadian businesses at their request. However, these technologies were not explicitly mentioned by our participants. It remains unclear to us whether these technologies are not being utilized by the businesses in our survey, or if the participants themselves are unaware of them. A future research could help identify where/whether the state-of-the-art PETs are being deployed and clarify their adoption across industries. This would provide valuable insights for both researchers and industry practitioners. 


\paratitle{Compliance encourages PETs adoption, but the verification of how PETs helps is unclear} We received several responses indicating the regulatory requirements encourage PETs adoption and mandate compliance for the adopted technologies, yet they lack details on how it is decided whether a PET can help with the regulation or is compliant with it, unless the technology is already a  ``best practice''. A deeper dive into this could be insightful for PETs designers. 

\paratitle{Investigations on how best practices emerge}``Best practices'' were repeatedly mentioned in our survey as a key resource for addressing various challenges, from regulatory compliance to selecting appropriate PETs solutions. Identifying what drives the initial emergence and integration of these best practices is essential for understanding how privacy technologies gain momentum and eventually become standard in industry practices. 

\paratitle{The effect of AI on privacy focus areas in industry} 
Future research on the effect of AI on privacy critical areas is crucial as AI technologies continue to evolve and integrate into various sectors. This effect is twofold. On one hand, it can complicate privacy compliance. For example, the common strategy of ``removing identifiers'' becomes even less effective in the age of AI, as its advanced pattern recognition and memorization capabilities~\cite{10.1145/3357713.3384290} enables re-identification more easily than ever before. Similarly, consent tracking or data disposal could become more challenging when the  data is used in model training. On the other hand, AI-driven solutions may assist with compliance; synthetic data produced by generative models that can replace the original sensitive data is one such example.

\section{Conclusion}\label{sect:conclusion}
Through this study, we identify the breadth of approaches employed by organizations considering PETs and the challenges they face. 
We further identify a gap between how companies think of privacy technologies and how researchers think of privacy technologies that can contribute to low adoption of the increasingly sophisticated privacy technologies produced by researchers, such as applications of differential privacy, multiparty computation, and trusted execution environments. 

Going forward, regulatory changes will continue to leave businesses without clarity as to how to meet the new requirements they create. 
Similarly, regulators will find it difficult to enforce these requirements without transparency into the technical systems designed with privacy protections in mind. Thus, clear communication between all parties is needed to ensure all parties share the same understanding of what technical privacy systems can and cannot guarantee. 
While emergent PETs may not have a clear adoption process as of yet, researchers and policymakers can improve the privacy of the populace through tailoring their efforts to better account for the needs of industry we have brought to the forefront with this work.

\section*{Acknowledgments}
The work of Xi He was supported by NSERC through a Discovery Grant, an alliance grant, and the Canada CIFAR AI Chairs program. The work of Bailey Kacsmar was supported by NSERC through a Discovery Grant (RGPIN-2024-04996).
\bibliographystyle{ACM-Reference-Format}


\appendix

\section{Survey Questions}\label{app:survQuestions}
The following are our six open-ended free-form text response questions we used as our survey. These were made available to our participants after the consent page. 
The order was not randomized and each participant received the exact same set of prompts. 

The informed consent page that was shown requested participants consent or decline to participate in the study. This consent page was also when they were informed about being able to skip questions, quit at any time, and any other relevant information for their decision to proceed through our study. 

\begin{enumerate}
    \item[SQ1] What are the current data handling best practices regarding potentially sensitive customer or client data within your company? This includes methods for collection, storage, analysis, and utilization of such data.
    \item[SQ2] When new privacy regulations (such as GDPR, CCPA, PIPEDA, or Bill C-27) emerge, what resources or plans does your company consider in determining the best course of action? 
    \item[SQ3] What processes does your organization employ to ensure compliance with existing or emerging privacy regulations?
    \item[SQ4] Please outline the decision-making process within your company regarding the adoption or development of privacy technologies. For example, when considering the adoption of methods like two-factor authentication.
    \item[SQ5] Please describe an example of a privacy-preserving technology method for the collection, storage, analysis, and utilization of sensitive data  that you think is (relatively) widely adopted in your domain of industry. 
    \item[SQ6] What factors facilitate the integration or creation of privacy technology methods for the collection, storage, analysis, and utilization of sensitive data within your domain of industry?
\end{enumerate}

\end{document}